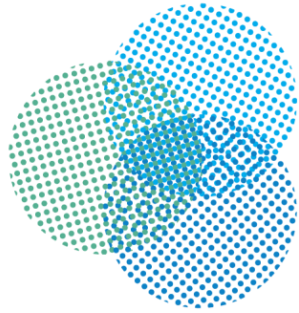

# Measuring and modeling the perception of natural and unconstrained gaze in humans and machines

Daniel Harari*, Tao Gao*, Nancy Kanwisher, Joshua Tenenbaum, Shimon Ullman

## Abstract

Humans are remarkably adept at interpreting the gaze direction of other individuals in their surroundings. This skill is at the core of the ability to engage in joint visual attention, which is essential for establishing social interactions. How accurate are humans in determining the gaze direction of others in lifelike scenes, when they can move their heads and eyes freely, and what are the sources of information for the underlying perceptual processes? These questions pose a challenge from both empirical and computational perspectives, due to the complexity of the visual input in real-life situations. Here we measure empirically human accuracy in perceiving the gaze direction of others in lifelike scenes, and study computationally the sources of information and representations underlying this cognitive capacity. We show that humans perform better in face-to-face conditions compared with 'recorded' conditions, and that this advantage is not due to the availability of input dynamics. We further show that humans are still performing well when only the eyes-region is visible, rather than the whole face. We develop a computational model, which replicates the pattern of human performance, including the finding that the eyes-region contains on its own, the required information for estimating both head orientation and direction of gaze. Consistent with neurophysiological findings on task-specific "face" regions in the brain, the learned computational representations reproduce perceptual effects such as the 'Wollaston illusion', when trained to estimate direction of gaze, but not when trained to recognize objects or faces.

* Daniel Harari and Tao Gao Contributed equally to this work.

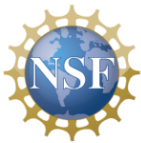

**This work was supported by the Center for Brains, Minds and Machines (CBMM), funded by NSF STC award CCF-1231216.**

# Introduction

Humans, as a social species, are remarkably adepts at understanding other people's mental states based on the perception of their actions (Blakemore & Decety, 2001; Carruthers & Smith, 1996). Developmental studies have demonstrated that even young infants can be engaged in joint attention with other humans (Yu & Smith, 2013) and can understand their mental states by observing and interpreting their non-verbal behavior, including their looking direction (Baron-Cohen, 1994; Johnson, Slaughter, & Carey, 1998; Chris Moore & Corkum, 1994; Saxe, Carey, & Kanwisher, 2004; Striano & Reid, 2006). Inferring where people are looking at plays a major role in the development of communication and language (Brooks & Meltzoff, 2005; Tomasello, 1995, 1999), opens a window into their mental state, and serves as an important cue towards understanding intentions and actions in social interactions (Fig. 1, (Calder et al., 2002; Ross Flom, Lee, & Muir, 2007; Frischen, Bayliss, & Tipper, 2007; C. Moore & Dunham, 1995; Mundy & Newell, 2007; D. Perrett & Emery, 1994; Reid & Striano, 2005; Scaife & Bruner, 1975; Vertegaal, Veer, & Vons, 2000)).

While gaze perception has been extensively studied in human vision (Bock, Dicke, & Thier, 2008; Cline, 1967; Gibson & Pick, 1963; Schweinberger, Kloth, & Jenkins, 2007; Stiel, Clifford, & Mareschal, 2014; Symons, Lee, Cedrone, & Nishimura, 2004; Todorović, 2006), the scope of existing behavioral studies provides only a limited understanding of gaze perception in real scenes. The main reason is that many behavioral studies focused on subjects' ability to judge whether or not a person's gaze was directed at them (Anstis, Mayhew, & Morley, 1969; Cline, 1967; Gibson & Pick, 1963; Schweinberger et al., 2007; Stiel et al., 2014). The acuity of detecting direct eye-contact was found to be as high as 3° of visual angle (Cline, 1967; Gibson & Pick, 1963). However, brain studies have shown that direct gaze and general gaze directions are encoded in different brain regions, suggesting that direct eye-contact may involve a separate processing mechanism and different computational representations compared with judging general direction of gaze (Allison, Puce, & McCarthy, 2000; Calder et al., 2002, 2007; George, Driver, & Dolan, 2001; Hoffman & Haxby, 2000; Jellema, Baker, Wicker, & Perrett, 2000; Wicker, Michel, Henaff, & Decety, 1998). Another limitation in past studies is due to the constraints, imposed by most of the behavioral studies, on the looker's gaze behavior, such as fixing the head at a certain pose and only allowing the eyes rotation (Bock et al., 2008; Symons et al., 2004). While these constraints help to isolate the effects of head and eye orientations (Kluttz, Mayes, West, & Kerby, 2009; S. R. Langton, 2000; S. R. H. Langton, Honeyman, & Tessler, 2004), it remains unclear from past studies, what is the accuracy of gaze perception under natural unconstrained looking , where the looker can move her head and gaze freely (Carpenter, 1988; Freedman & Sparks, 1997), and which parts of the face contribute most of the information required to perform the task. Dealing with unconstrained gaze direction is also a more challenging computational problem. In frontal-view with controlled head movement, the eyes are clearly visible to the observer, and the gaze direction can be estimated directly from the relative position of the iris and pupil in the eyes. In contrast, in the unconstrained gaze scenario, the appearance of the eyes vary dramatically, from clear front-view of the eye, to partially occulated eyes, when most of the eyes are invisible, due to the complex interaction of head pose and eye gaze. Finally, in many studies on discriminating gaze perception, the looker is asked to look at an empty space (Cline, 1967; Gibson & Pick, 1963), whereas in natural and realistic situations a person's gaze is typically oriented at an object in the scene. Both developmental and neuropsychological studies have shown that gazing towards an object is an important source for learning and estimating direction of gaze (Amano, Kezuka, & Yamamoto, 2004; D'Entremont, Hains, & Muir, 1997; Ross Flom, Deák, Phill, & Pick, 2004; Pelphrey, Singerman, Allison, & McCarthy, 2003; Ullman, Harari, & Dorfman, 2012), and it remains possible that the experimental conditions in the 'empty space' studies, were not identical to object-directed gaze.

Computationally, modeling the perceptual and cognitive processes involved in the analysis of social interactions, including the extraction of 3D direction of gaze for interpreting visual joint attention in real-life situations, pose a challenge due to the complexity of the visual input in realistic 3D environment, compared with restricted 3D configurations studied under laboratory conditions. Head pose estimation and detection of eye gaze have been studied extensively in the computer vision and applied mathematics communities (Funes Mora, Nguyen, Gatica-Perez, & Odobez, 2013; Gee & Cipolla, 1994; Hansen & Ji, 2010; Lu, Sugano, Okabe, & Sato, 2011; Mora & Odobez, 2012; Murphy-Chutorian & Trivedi, 2009; Odobez & Mora, 2013; Recasens, Khosla, Vondrick, & Torralba, 2015; Sugano, Matsushita, & Sato, 2014; Todorović, 2006; Weidenbacher, Layher, Bayerl, & Neumann, 2006; Wood et al., 2015; Zhang, Sugano, Fritz, & Bulling, 2015). The majority of these studies addressed either the head pose estimation or the eye gaze direction estimation as disjoint problems (Gee & Cipolla, 1994; Krüger, Pötzsch, & von der Malsburg, 1997; Lu et al., 2011; Murphy-Chutorian & Trivedi, 2009; Odobez & Mora, 2013; Wang, Wang, Yin, & Kong, 2003). While studies of these two problems have demonstrated impressive performance for each of the separate tasks, little has been done in addressing the problem of detecting the direction of gaze in natural and unconstrained scenes, in which observed humans can look freely at different targets. In this natural and unconstrained setting, gaze events towards different targets may share the same head or eye poses, while different poses may share the gaze towards the same target. A recent study (Recasens et al., 2015) trained a deep neural network to infer direction of gaze in natural images using many labeled examples, in which both head orientation and gaze targets were manually annotated. Since heads and faces in the training set images were small and sometimes seen from the back, it remains unclear which face and eye parts contributed to the gaze estimation. Other studies (Odobez & Mora, 2013; Zhang et al., 2015) suggested to estimate direction of gaze under free-head movements by first rectifying the eyes images (estimating how they would look if seen from a frontal view), using the 3D head pose, acquired by a 3D sensor or estimated from the face image. However, the eyes rectification procedure is complex and sets limitation on the range of supported head poses (due to self-occlusions at non-frontal poses), and there is no evidence for its role in human perception. Eye-region rectification was addressed in another study (Wood et al., 2015), which suggested a computer graphics method for synthesizing realistic close-up images of the human eye for a wide range of head poses, gaze directions, and illumination conditions. Using the synthesized eye images in the training phase improved the performance over state-of-the-are methods for gaze estimation including deep neural nets, but the synthesized dataset was designed for a typical laptop-viewing setting with limited head pose and gaze variations.

Here we study for the first time accuracy of 3D gaze perception under natural and unconstraint looking conditions. In our settings, an 'observer' interprets the 3D gaze direction of a human 'looker', which can move her head and eyes freely, looking at objects around her (Fig. 2). Several conditions were tested to determine the source of information for the estimation of the 3D gaze direction: 'live' versus 'recorded' stimuli, 'dynamic' versus 'static' and 'whole-face' visibility versus 'eyes-region' only and 'face-without-eyes'. The results show that humans perform better in the 'live' condition compared with the 'recorded' condition, but not due to the dynamics of the visual input. Furthermore, the eyes-region (Fig. 3) was essentially sufficient for estimating 3D direction of gaze and yields equivalent performance to the 'whole-face' condition. To better understand human performance, we constructed and compared computational models that can process head orientation and gaze direction from 2D images of faces. We developed a model that replicates human performance under similar 'recorded' conditions (leaving 'live' conditions to future studies), including the finding that the eyes region contains on its own most of the information for interpreting 3D direction of gaze and head orientation (Emery, 2000; S. R. Langton, 2000; D. I. Perrett, Hietanen, Oram, Benson, & Rolls, 1992; Todorović, 2006; Wollaston, 1824). The model operates in a two-stage process: head pose estimation followed by gaze direction estimation from the eyes 'conditioned' on head pose. The two-stage processing proved superior to end-to-end deep neural networks (DNN) we studied. In addition, the learned representations show an effect similar to the 'Wollaston illusion' (Fig. 4, (Wollaston, 1824)) when trained for estimating direction of gaze but not for object or face recognition, suggesting the involvement of gaze-selective cortical regions in the Wollaston effect.

## Methods

This study combines empirical testing with computational modeling and evaluation. The goal was to measure human accuracy, and unfold the sources of information, which underlie the perceptual and cognitive processes involved. We first describe the empirical studies, conducted under different conditions. Next, we describe the computational study, which includes modeling of the computational processes, and a comparison between the performance of models and humans under similar conditions.

### *Human Experiments*

Tasks:

Each trial of the current study involved two human participants: a looker and an observer. The looker was sitting, facing 52 objects arranged on a tabletop. On each trial, a command was generated for the looker, indicating the color and number of a target object. The looker was instructed to look at the target object "as naturally as possible" by moving the head and eyes freely. An observer, either sitting in front of the looker across the table, or watching a recorded session of the looker, saw the looker's gazing action as well as all objects laid on the table.
The task for the observer was to infer the target object of the looker's gaze. In each block of trials, each object was selected as the target once in a random order generated by an automated script. The precision of the observer's response was the primary psychophysical measurement, which was also used in the evaluation of the computational study. All participants were with normal or corrected-to-normal vision, gave their informed consent and were paid for their participation. All experiments and procedures were approved by the institutional review board of Massachusetts Institute of Technology, Cambridge, MA, USA.

Experimental Design:

Across two experiments, we varied the visual input of the observers, to control the sources of visual information available to them in performing the task. As described in more detail below, the viewing conditions ranged from fully unconstrained live 3D observation, down to highly blurred face images with two tightly constrained patches allowing a clear view only of the eyes. This design enabled us not only to measure the overall precision of gaze perception, but also to reveal the contribution of different sources of information, such as 3D vs. 2D, dynamic vs. static and clear view of the whole face vs. partially blurred views of the face.

### *Experiment 1a and 1b: Live conditions*

We started from the viewing condition that contained the richest visual information: a face-to-face live 3D viewing (*LC*). Four observers sat at the opposite side of the table watching a single looker. The looker looked at target objects, one target in each trial, by following commands showing on a monitor (visible only to the looker, Figure S1A). At the beginning of the experiment, the looker was instructed to keep looking at the target, until a beep sound signaled the end of the trial after 10 seconds, when the looker should look back at the monitor for the command of the next trial. The observers were instructed to write down the name of the target object by interpreting the lookers gaze. Following each two blocks, the observers shift their sits clockwise. The four positions were categorized into two classes: center and periphery. To make a direct comparison with results from the recorded conditions in experiment 2 (the recording equipment was located in the center), here we only include in the results data from the two center positions (Figure S1B).

Experiment 1a has 2 Asian lookers, one female (JP) and one male (TG). There were 16 Asian participants as the observers. Experiment 2b has two Caucasian lookers, one female (VO) and one male (HE). There were 10 Caucasian and African American participants as the observers in this part of the experiment. Experiment 1a originally also included a sunglasses condition, in which the looker wears a pair of dark sunglasses that covers the eyes region. The observers' performance in this condition significantly drops. However, wearing sunglasses disrupts the automatic face tracking of the recording system, which makes the data unavailable for computational modeling. Therefore, we do not include data from this condition in the results and exclude it from experiment 2b.

*Experiment 2a and 2b: Recorded conditions*

Here we showed the observers the recorded videos and images (captured by the RGB-D Kinect sensor) of the lookers' actions on a computer screen. The videos and images show clearly the whole scene, including both the looker and the full object array on the table (the Kinects field of view is 70º×60º of visual angle). The face of the looker roughly spans 2º-3º of visual angle in the observer's field of view (compares to roughly 5º-6º of visual angle in the live condition). Although 3D live information is no longer available to the observers, we can systematically manipulate the visual stimuli, by introducing 4 different viewing conditions in Exp. 2a (*Supplementary Material*): (*RC1*) Dynamic video: the entire movements of the looker were shown to the observer, starting from the resting state (watching a command on the screen behind the recording system) and ending at the fixed gaze at the target; (*RC2*) Whole-face static images: Images of lookers gazing at targets were shown. Image frames were extracted from the recorded videos when the lookers kept steady both their head and eyes (Fig. S3A); (*RC3*) Eyes-region static images: a rectangular image strip around the eyes region (including the nose bridge and face boundaries at the temples) was visible, while the rest of the face was blurred (using a low-pass Gaussian filter, Fig. S3B); (*RC4*) Head-only static images: The eyes-region was Gaussian-blurred, while the rest of the face was visible (Fig. S3E).

Exp. 2b further isolated information available in the eyes-region by introducing two further conditions: (*RC5*) Tight-eyes: face parts and boundaries surrounding the eyes (excluding the nose bridge) were Gaussian-blurred (Fig. S3C). (*RC6*) Separate-eyes: the nose bridge between the two eyes was also Gaussian-blurred (Fig. S3D). The Whole-face (*RC2*) and Eyes-region (*RC3*) conditions were repeated in Expt. 2b as baselines.

Images of the Eyes-region, Head-only, Separate-eyes and Tight-eyes conditions (*RC3-RC6*) were automatically created using standard computer vision algorithms (*Supplementary Material*).

In both experiments 2a and 2b, there were 16 blocks (4 visual conditions for each of the four lookers). Each block consisted of 52 trials, in which the order of the target objects was randomized. Each trial lasted 10 seconds. Observers were requested to move a computer mouse and click on the inferred target. Both block order and trial order within each block, were randomized. Each observer had 10 practice trials at the beginning of the experiment, using images of a different looker which was not one of the four lookers of the actual experiment.

Apparatus and stimuli:

An array of 52 objects (candles of size 4.6×4.6×3.6 cm) was laid on a table in a concentric configuration, with 13 columns and four rows (semi-rings, Fig. S1B). A red wooden egg was placed on the table to mark the center position of the concentric array. Objects on the same row had the same color (white, green, blue or red). The number of the column was marked on the side of each object. The distance of each row from the center of the array (29 cm for the closest row, and 96 cm for the furthest row) was set such that the visual angle between every two adjacent rows, from a point 35 cm above the center of the array (the average looker perspective at the array), was 10°. The column positions were set such that the angular difference between each two adjacent columns was 10° on the table surface. The corresponding visual angle between every two adjacent columns, from the same point (35 cm above the center of the array), was 8° on average. In practice, the visual angles slightly vary given the exact position of the looker's head and were computed on a trial-by-trial basis (mean difference of viewing angle between lookers across all target objects was 4.7º±3.0º). A Microsoft Kinect V2 RGB-D sensor was also positioned across the table, facing the table and the object array (121 cm away from the center of the array and slightly below the instruction screen, Fig S1A). The RGB-D sensor was used to record on video the looking trials of the lookers, and also provide accurate 3D information of the recorded scene for computational evaluations (RGB at 1920×1080 pixel resolution; depth at 512×424 pixel resolution; see (Sarbolandi, Lefloch, & Kolb, 2015) for a detailed evaluation of the Kinect's depth accuracy).

In the live condition, observers were seated on the opposite side of the table, 128cm away from the looker. In the recorded conditions, observers were seated 60 cm away from the display (full screen size is 30° × 19° of visual angles). The recorded stimuli included 4 video sessions and 1040 still images of looking trials at different viewing conditions, all at 1920×1080 pixel resolution (Fig. S2, S3, *Supplementary material*).

### *Computational modeling*

To understand the sources of information for the underlying perceptual and cognitive processes in estimating a person's 3D gaze direction, we studied computational models for recognizing 3D direction of gaze from 2D images of faces. Accurate 3D information of the visual environment was extracted from the RGB-D sensor depth data, including the 3D position of objects and faces. The Microsoft Kinect for Windows SDK provides accurate face tracking and reliable 3D head orientation for tracked faces. The 3D direction of gaze was defined as the 3D vector pointing from the center location between the eyes and the location of the target object.

We developed a model for estimating the 3D direction of gaze from a 2D image of a face or face parts, based on state-of-the-art computer vision methods. The model works in a two-stage process, head pose first, and then gaze direction from eyes 'conditioned' on the head pose. In our implementation image representations of the face and face parts (Bosch, Zisserman, & Munoz, 2006; Dalal & Triggs, 2005; Lowe, 2004) are associated with 3D directions of the head orientation and gaze direction (Altmann, 1986; Wilkins, 1844), using the $k$-nearest neighbors ($k$-NN) approach (Duda, Hart, & Stork, 2001; Wu et al., 2008) similar to ((Ullman et al., 2012), $k$=15, *Supplementary material*).

### *Computational evaluation*

A dataset for training and evaluation was created from recorded sessions of 10 'lookers', which were not used in the testing of the human 'observers'. The dataset consists of 1916 face images and their associated 3D head orientation (as extracted from the Kinect's face tracking algorithm) and 3D direction of gaze (between the location of the face and instructed target object), collected from 37 blocks. Evaluation on this dataset was done using a leave-one-looker-out cross-validation approach.

We evaluated our two stage model together with leading deep neural network (DNN) models (Krizhevsky, Sutskever, & Hinton, 2012; LeCun, Bengio, & Hinton, 2015), that were pre-trained for object recognition (Simonyan & Zisserman, 2015) or face recognition (Parkhi et al., 2015), and then fine-tuned to infer 3D head orientation and 3D direction of gaze from 2D face images (the original '*softmax*' output classification layers were replaced with '*fully connected*' output layers of size 4, representing 3D directions in rotation quaternions). Models were evaluated on similar conditions to the human psychophysics, excluding the dynamic condition (*RC1*). For comparison with human performance, the models were also tested on face images extracted from the stimuli images of the human psychophysics test (*RC2-RC6*).

To evaluate the Wollaston illusion (Wollaston, 1824), a set of 120 face triplets was created from the evaluation dataset. A face triplet consisted of a pair of authentic face images (of the same 'looker'), and an artificially 'synthesized' face image (Fig. 4). The pair of authentic face images depicted a face turned to the left in one image and to the right in the other image (mean angular difference between the head orientations in the two images M=25.9º, SD=7.6º, and between the gaze directions M=40.3º, SD=8.0º). An image cloning method (Pérez, Gangnet, & Blake, 2003; Tanaka, Kamio, & Okutomi, 2012) was used to generate a synthesized face image with the same eyes from the first image ('source') and head pose from the second image ('target'), by replacing the eyes in the 'target' image with the cropped eyes from the 'source' image. We applied the DNN models to the set of face triplets and compared between the estimated 3D directions of gaze. We also compared the underlying representations for the eyes, which were extracted from the deepest '*pooling*' layer of the models, in models trained for gaze estimation, as well as models trained for object and face recognition. For the comparison, we used correlations of the extracted layer responses localized at the eyes region, between pairs of faces with different head poses and either different or same eyes.

## Results

### *Human experiments results*

Figure 5A shows the overall performance of human observers in estimating the 3D gaze direction of lookers under the tested conditions (*Methods LC, RC1-RC6*). The accuracy is measured as the percentage of trials in which the human observers get exactly the correct target ('exact' accuracy in red; mean angular error (MAE) is less than 4.7º), or one object off the correct target ('4-neighborhood'[1] accuracy in green, '8-neighborhood'[2] accuracy in blue; MAE is less than 14.1º and 17.0º, respectively). The 'exact' accuracy in the 'live' condition (M=46%, SD=13%, *Methods LC*) was significantly higher (t(31)=2.890, p=0.007) than the highest 'exact' accuracy among the tested 'recorded' conditions (dynamic video, M=31%, SD=6%, *Methods RC1*). Both the 4- and 8-neighborhood accuracies of the 'live' condition were significantly higher as well (4-n: t(31)=5.244, p<0.001; 8-n: t(31)=6.730, p<0.001). However, the difference between the 'exact' accuracy in the dynamic condition and the 'exact' accuracy in the static whole-face condition (M=28%, SD=7%, *Methods RC2*) was found to be insignificant (t(6)=1.474, p=0.191), suggesting that the performance gap between the 'live' and 'recorded' conditions is not due the input dynamics. The 'exact' accuracy in the 'eyes-region' condition (M=21%, SD=2%, *Methods RC3*) was significantly less than the 'exact' accuracy in the whole-face condition (t(6)=3.60, p=0.011, *Methods RC2*), but the difference between the 4- and 8-neighborhood accuracies of the two conditions was not significant (4-n: t(6)= 1.494, p=0.186; 8-n: t(6)= 1.757, p=0.130). Furthermore, both 'exact', 4- and 8-neighborhood accuracies in the eyes-region condition were much better than the accuracies in the head-only condition ('exact': M=11%, SD=3%, t(6)=7.749, p<0.001; 4-n: t(6)= 6.313, p<0.001; 8-n: t(6)= 5.241, p=0.002; *Methods RC4*). These findings suggest that the 'eyes-strip' essentially contains most of the information needed for accurate 3D gaze estimation.

In Experiment 2b, we furthered manipulated images around the eyes-region. Removing the bridge of the two eyes were most effective, as 'exact' accuracy of the separate-eye condition (M=19%, SD=4%, *Methods RC6*) was significantly lower than that of the eyes-region condition (t(7)=3.677, p=0.008). 'Exact' accuracy of the tight-eyes condition (M=21%, SD=4%, *Methods RC5*) was also lower than that of the eyes-region condition, with a marginal significance (t(7)=2.315, p=0.054).

Remarkably, the human accuracy in all conditions is doubled in the '4-neighborhood' analysis and tripled in the '8-neighborhood', while the MAE is less than 17º.

### *Results for the computational model*

For the computational model evaluation, accuracy was measured directly as the angular error between the estimated 3D direction and the true 3D direction of both head orientation and gaze to target. To compare between the model's accuracy and the human accuracy in each trial, we also extracted an estimated target object, as the nearest object around the intersection between the estimated 3D direction of gaze and the 3D plane of the object array. Applying the model to the same test images from the human experiments, as well as on additional evaluation images, reproduced the performance variations under similar conditions (*Methods RC2-RC6,* Fig. 5B), although humans performed better than the computational model. As with humans, the accuracy in the 'eyes-region' only condition (MAE=9.9º, SD=6.7º, 'exact' accuracy=18%, *Methods RC3*) was found to be comparable with the accuracy in the whole face condition (MAE=9.7º, SD=5.5º, 'exact' accuracy=19%, *Methods RC2*). Evaluation of the estimated head orientation during the first stage of our model, yielded comparable mean angular errors when applied to the 'eyes-region' (MAE=9.3º, SD=5.8º) and to the whole face (MAE=6.9º, SD=4.8º), suggesting that the 'eyes-region' essentially contains full head pose information (mainly extracted from the face boundaries near the temples and the nose bridge, Table S1). Alternative single-stage models, including deep neural networks, which estimate directly both head orientation and gaze direction from a given face image, were trained and evaluated, but were all found to be inferior to the two-stage model (*Supplementary Material*).

### *Computational results on the Wollaston illusion*

The two-stage model inherently supports the Wollaston illusion that the perceived gaze of identical eyes with different head-pose contexts is shifted in the direction of the head orientation, since in the model the eyes are conditioned on the head orientation. For a quantitative analysis of the underlying representations of the eyes in the model, we evaluated a deep neural network (DNN) trained for gaze estimation (*Methods, Computational evaluation*). This DNN reproduced the Wollaston illusion when applied to the 'synthesized' face images in our dataset, with mean angular offset of M=31.8º, SD=8.5º, from the gaze direction in the 'source' images with identical eyes but different head orientation (the estimated gaze direction in the 'synthesized' images was in the range of [-9.5º, +9.5º] around the 'direct' gaze towards the camera). These results suggest that the eyes representations for congruent and in-congruent eyes, with respect to the head orientation as implied by the face context, are different. We further evaluated networks trained for object classification and face identification (*Methods, Computational evaluation*), which yielded with significantly lower correlation values between network responses to the eyes region in two authentic face images ('source' and 'target' with dissimilar eyes), than the correlation values between the responses to the eyes region in 'source' and 'synthesized' images, in which the eyes are identical (t(238)=-4.00, p<0.001, for object classification; t(238)=-1.66, p=0.05, for face identification). The lower correlation values suggest that these networks have similar representations for the identical eyes in the 'source' and 'synthesized' images, but different representations for the dissimilar eyes in the 'source' and 'target' images. However, for the network trained for gaze estimation, the difference between the correlation values for the two pairs of images was found to be insignificant (t(238)=-1.07, p=0.14), suggesting that in this network the representations for the identical eyes in the 'source' and 'synthesized' images, are as different as the representations for the dissimilar eyes in the 'source' and 'target' images. This is related to Wollaston comment that for humans, identical eyes look different with different head-pose contexts.

## Discussion

The current study aims to better understand the perception of 3D gaze direction under natural and unconstraint looking conditions at a single target out of tens of objects. Studying unconstrained gaze direction is a challenging computational problem, due to the complex interaction between head orientation and eye gaze direction. The gaze direction cannot be estimated directly from the relative position of the pupils in the eyes, as in the front-view case. Furthermore, the appearance of the eyes varies dramatically, from fully visible front-view of the two eyes, to partially occulated eyes, where most of the eyes region becomes invisible.

In this study, we have measured for the first time human accuracy of perceiving unconstrained natural gaze directions of human lookers. The experiments tested several viewing conditions, including face-to-face (live) and recorded conditions. A comparison between the accuracy of the different conditions indicates that the performance in the live condition is significantly better than in the recorded conditions. Further analysis shows that the gap is not due to the dynamics in the input, similar to the findings in (Symons et al., 2004). It is worth noting, however, that input dynamics serve as a strong learning cue for gaze following in early infanthood, with static gaze perception developing only later, around the age of 12 months (Brooks & Meltzoff, 2005; D'Entremont et al., 1997; Meltzoff & Brooks, 2007; Ullman et al., 2012). The performance gap between live and recorded conditions may be in part due to better 3D perception using e.g. stereoscopic vision in the live condition, as indicated by a recent study (which was limited to frontal views, (Atabaki, Marciniak, Dicke, & Thier, 2015)). This assumption may be further validated empirically, using for instance monocular vision by covering one eye. The gap may be also due to body context cues, which may be better perceived during the live condition, and were found to influence face and gaze perception (Moors, Germeys, Pomianowska, & Verfaillie, 2015; D. I. Perrett et al., 1992; Yovel, Pelc, & Lubetzky, 2010).

Among the static viewing conditions, the highest accuracy was measured when the whole face was visible in the stimuli, as expected. However, surprisingly, the small performance gap between viewing the whole face compared with the eyes region only, suggests that the eyes region includes on its own most of the information required for accurate perception of the 3D direction of gaze. This was unexpected, as this limited region contains only a fraction of the face features. Furthermore, our computational model, which replicates these findings, can also extract head orientation at human-level accuracy based on the eyes region only. Excluding the face boundaries and the nose bridge from the eyes-region stimuli results in a significant performance drop for both humans and model, since these face parts provide essential information, in particular for 3D head pose estimation (S. R. H. Langton et al., 2004). Having a compact region, where most of the relevant information is located, may provide an efficient computational representation for gaze estimation, which can be acquired in a single saccade, when the entire eyes region falls within foveal vision. It will be interesting to test the patterns of fixations on faces, while performing e.g. the task of gaze estimation during triadic joint attention, and compare with patterns from other face-related tasks (Peterson & Eckstein, 2013).

Our computational model consists of two processing stages: estimation of head orientation, followed by estimation of gaze direction from the eyes, 'conditioned' on the estimated head pose. Moreover, both attraction and repulsion effects of the head orientation on the perceived gaze are implicitly included in the model (Otsuka, Mareschal, Calder, & Clifford, 2014). The model proves to be superior to alternative single-stage models, including end-to-end deep neural network models, which learn to associate images of face and facial parts directly to gaze direction and head orientation. Performance evaluation of our computational model reproduces human performance, including the differences among the different viewing conditions. The model also agrees with neurophysiological findings on the organization of face processing cells for social attention in the STS, that are responsive for head pose and gaze in the same direction(Calder et al., 2007; De Souza, Eifuku, Tamura, Nishijo, & Ono, 2005; Farzmahdi, Rajaei, Ghodrati, Ebrahimpour, & Khaligh-Razavi, 2016; Freiwald & Tsao, 2010; D. I. Perrett et al., 1992). The two-stage model, is also in agreement with psychophysical findings on the development of face and gaze perception, as infants are first able to track faces and roughly follow gaze using head motion (D'Entremont et al., 1997), while using cues from the eyes for accurate gaze perception only at later stages (Ross Flom et al., 2004; Johnson et al., 1998; Meltzoff & Brooks, 2007). This developmental trajectory may suggest that the computational processing of the eyes for gaze perception is built on top of (and therefore conditioned on) the early acquired capability of head pose estimation.

An intriguing finding is the reproduction of the Wollaston illusion (both for perceived gaze and eyes appearance) with a model trained for estimating gaze direction, but not with alternative models trained for object classification or face recognition. This finding suggests that different face and eyes representations are learned in computational DNN-based models for different tasks, in line with neurophysiological findings on task-specific cortical regions, which are responsive to face and facial parts, including the eyes, for gaze estimation, but not for face or object recognition (Calder et al., 2007; Carlin, Calder, Kriegeskorte, Nili, & Rowe, 2011; Hoffman & Haxby, 2000). Our findings suggest that the Wollaston effect depends specifically on the gaze-related regions.

Finally, our computational model for accurate estimation of 3D gaze direction from face images could be combined with existing methods for depth estimation (Liu, Shen, Lin, & Reid, 2015) and scene segmentation (Shi & Malik, 2000), to model joint attention in social interactions. In particular, such a combined scheme will be able to follow direction of gaze in 3-D space and identify the attended target. An artificial intelligence system, which includes the cognitive capability for interpreting joint attention, will be able to interpret social interactions in scenes, by understanding that some people in the scene are engage in joint attention, as well as learn to interact on its own (e.g. a robot) in social interactions by identifying attended targets of humans in its view (Fig. 6).

## Notes

1. 4-neighborhood, also known as the Von Neumann neighborhood, comprises the four objects orthogonally surrounding a central target in the object array. Mathematically, it is defined as the set of points at a Manhattan distance of 1 from the central point.

2. 8-neighborhood, also known as the Moore neighborhood, comprises the eight objects surrounding a central target in the object array. Mathematically, it is defined as the set of points at a Chebyshev distance of 1 from the central point.

## References

Allison, T., Puce, A., & McCarthy, G. (2000). Social perception from visual cues: Role of the STS region. *Trends in Cognitive Sciences*, *4*(7), 267–278. http://doi.org/10.1016/S1364-6613(00)01501-1

Altmann, S. L. (1986). *Rotations, Quaternions, and Double Groups*. Dover Publications. http://doi.org/10.1002/qua.560320310

Amano, S., Kezuka, E., & Yamamoto, A. (2004). Infant shifting attention from an adult's face to an adult's hand: a precursor of joint attention. *Infant Behav Dev*, *27*(1), 64–80.

Anstis, S., Mayhew, J., & Morley, T. (1969). The perception of where a face or television'portrait'is looking. *The American Journal of Psychology*.

Atabaki, A., Marciniak, K., Dicke, P. W., & Thier, P. (2015). Assessing the precision of gaze following using a stereoscopic 3D virtual reality setting. *Vision Research*, *112*, 68–82. http://doi.org/10.1016/j.visres.2015.04.015

Baron-Cohen, S. (1994). How to build a baby that can read minds: Cognitive mechanisms in mindreading. *Cahiers de Psychologie Cognitive/Current Psychology of Cognition*, *13*, 513–552. article.

Blakemore, S.-J., & Decety, J. (2001). From the Perception of Action to the Understanding of Intention. *Nature Reviews Neuroscience*, *2*(August), 561–567. http://doi.org/10.1038/35086023

Bock, S. W., Dicke, P., & Thier, P. (2008). How precise is gaze following in humans? *Vision Research*, *48*(7), 946–57. http://doi.org/10.1016/j.visres.2008.01.011

Bosch, A., Zisserman, A., & Munoz, X. (2006). Scene classification via pLSA. In *Proc. 9th European Conference on Computer Vision* (Vol. 3954 LNCS, pp. 517–530). Springer. http://doi.org/10.1007/11744085_40

Brooks, R., & Meltzoff, A. N. (2005). The development of gaze following and its relation to language. *Developmental Science*, *8*(6), 535–543. http://doi.org/10.1111/j.1467-7687.2005.00445.x

Calder, A. J., Beaver, J. D., Winston, J. S., Dolan, R. J., Jenkins, R., Eger, E., & Henson, R. N. a. (2007). Separate coding of different gaze directions in the superior temporal sulcus and inferior parietal lobule. *Current Biology : CB*, *17*(1), 20–5. http://doi.org/10.1016/j.cub.2006.10.052

Calder, A. J., Lawrence, A. D., Keane, J., Scott, S. K., Owen, A. M., Christoffels, I., & Young, A. W. (2002). Reading the mind from eye gaze. *Neuropsychologia*, *40*(8), 1129–1138. http://doi.org/10.1016/S0028-3932(02)00008-8

Carlin, J. D., Calder, A. J., Kriegeskorte, N., Nili, H., & Rowe, J. B. (2011). A head view-invariant representation of gaze direction in anterior superior temporal sulcus. *Current Biology : CB*, *21*(21), 1817–21. http://doi.org/10.1016/j.cub.2011.09.025

Carpenter, R. H. S. (1988). *Movements of the Eyes*. London: Pion Limited.

Carruthers, P., & Smith, P. K. (Eds.). (1996). *Theories of Theories of Mind*. Cambridge: Cambridge University Press.

Cline, M. (1967). The perception of where a person is looking. *The American Journal of Psychology*.

D'Entremont, B., Hains, S. M. J., & Muir, D. W. (1997). A demonstration of gaze following in 3- to 6-month-olds. *Infant Behav Dev*, *20*(4), 569–572. http://doi.org/10.1016/S0163-6383(97)90048-5

Dalal, N., & Triggs, B. (2005). Histograms of Oriented Gradients for Human Detection. In *Proceedings of Computer Vision and Pattern Recognition* (pp. 886–893). CONF, Washington, DC, USA: IEEE Computer Society.

De Souza, W. C., Eifuku, S., Tamura, R., Nishijo, H., & Ono, T. (2005). Differential characteristics of face neuron responses within the anterior superior temporal sulcus of macaques. *Journal of Neurophysiology*, *94*(2), 1252–66. http://doi.org/10.1152/jn.00949.2004

Duda, R. O., Hart, P. E., & Stork, D. G. (2001). *Pattern classification*. John Wiley & Sons. http://doi.org/10.1038/npp.2011.9

Emery, N. J. (2000). The eyes have it: The neuroethology, function and evolution of social gaze. *Neuroscience and Biobehavioral Reviews*, *24*(6), 581–604. http://doi.org/10.1016/S0149-7634(00)00025-7

Farzmahdi, A., Rajaei, K., Ghodrati, M., Ebrahimpour, R., & Khaligh-Razavi, S.-M. (2016). A specialized face-processing model inspired by the organization of monkey face patches explains several face-specific phenomena observed in humans. *Scientific Reports*, *6*(April), 25025. http://doi.org/10.1038/srep25025

Flom, R., Deák, G. O., Phill, C. G., & Pick, A. D. (2004). Nine-month-olds' shared visual attention as a function of gesture and object location. *Infant Behavior and Development*, *27*(2), 181–194. http://doi.org/10.1016/j.infbeh.2003.09.007

Flom, R., Lee, K., & Muir, D. (Eds.). (2007). *Gaze-following: Its development and significance*. Mahwah, NJ: Lawrence Erlbaum Associates Publishers.

Freedman, E. G., & Sparks, D. L. (1997). Eye-head coordination during head-unrestrained gaze shifts in rhesus monkeys. *Journal of Neurophysiology*, *77*(5), 2328–2348.

Freiwald, W. A., & Tsao, D. Y. (2010). Functional Compartmentalization and Viewpoint Generalization Within the Macaque Face-Processing System. *Science*, *330*(November), 845–851. http://doi.org/10.1126/science.1229223

Frischen, A., Bayliss, A., & Tipper, S. (2007). Gaze cueing of attention: visual attention, social cognition, and individual differences. *Psychological Bulletin*, *133*(4), 694–724. http://doi.org/10.1037/0033-2909.133.4.694.Gaze

Funes Mora, K. a., Nguyen, L., Gatica-Perez, D., & Odobez, J.-M. (2013). A semi-automated system for accurate gaze coding in natural dyadic interactions. *Proceedings of the 15th ACM on International Conference on Multimodal Interaction - ICMI '13*, 87–90. http://doi.org/10.1145/2522848.2522884

Gee, A., & Cipolla, R. (1994). Determining the gaze of faces in images. *Image and Vision Computing*, *12*(10), 639–647. http://doi.org/10.1016/0262-8856(94)90039-6
George, N., Driver, J., & Dolan, R. J. (2001). Seen gaze-direction modulates fusiform activity and its coupling with other brain areas during face processing. *NeuroImage*, *13*(6 Pt 1), 1102–1112. http://doi.org/10.1006/nimg.2001.0769
Gibson, J. J., & Pick, A. D. (1963). Perception of another person's looking behavior. *The American Journal of Psychology*, *76*(3), 386–394. http://doi.org/10.2307/1419779
Hansen, D. W., & Ji, Q. (2010). In the eye of the beholder: a survey of models for eyes and gaze. *IEEE Transactions on Pattern Analysis and Machine Intelligence*, *32*(3), 478–500. http://doi.org/10.1109/TPAMI.2009.30
Hoffman, E. a, & Haxby, J. V. (2000). Distinct representations of eye gaze and identity in the distributed human neural system for face perception. *Nature Neuroscience*, *3*(1), 80–84. http://doi.org/10.1038/71152
Jellema, T., Baker, C. I., Wicker, B., & Perrett, D. I. (2000). Neural representation for the perception of the intentionality of actions. *Brain and Cognition*, *44*(2), 280–302. http://doi.org/10.1006/brcg.2000.1231
Johnson, S., Slaughter, V., & Carey, S. (1998). Whose gaze will infants follow? The elicitation of gaze-following in 12-month-olds. *Developmental Science*, *1*(2), 233–238. http://doi.org/10.1111/1467-7687.00036
Kluttz, N. L., Mayes, B. R., West, R. W., & Kerby, D. S. (2009). The effect of head turn on the perception of gaze. *Vision Research*, *49*(15), 1979–1993. http://doi.org/10.1016/j.visres.2009.05.013
Krizhevsky, A., Sutskever, I., & Hinton, G. (2012). Imagenet classification with deep convolutional neural networks. *Proc NIPS*, 1–9.
Krüger, N., Pötzsch, M., & von der Malsburg, C. (1997). Determination of face position and pose with a learned representation based on labelled graphs. *Image and Vision Computing*, *15*(8), 665–673. http://doi.org/10.1016/S0262-8856(97)00012-7
Langton, S. R. (2000). The mutual influence of gaze and head orientation in the analysis of social attention direction. *The Quarterly Journal of Experimental Psychology. A, Human Experimental Psychology*, *53*(3), 825–45. http://doi.org/10.1080/713755908
Langton, S. R. H., Honeyman, H., & Tessler, E. (2004). The influence of head contour and nose angle on the perception of eye-gaze direction. *Perception & Psychophysics*, *66*(5), 752–771. http://doi.org/10.3758/BF03194970
LeCun, Y., Bengio, Y., & Hinton, G. (2015). Deep learning. *Nature*, *521*(7553), 436–444. http://doi.org/10.1038/nature14539
Liu, F., Shen, C., Lin, G., & Reid, I. D. (2015). Learning Depth from Single Monocular Images Using Deep Convolutional Neural Fields. *PAMI*, 15. Retrieved from http://arxiv.org/abs/1502.7411
Lowe, D. G. (2004). Distinctive image features from scale-invariant keypoints. *International Journal of Computer Vision*, *60*(2), 91–110.
Lu, F., Sugano, Y., Okabe, T., & Sato, Y. (2011). Inferring human gaze from appearance via adaptive linear regression. *2011 International Conference on Computer Vision*, 153–160. http://doi.org/10.1109/ICCV.2011.6126237
Meltzoff, A., & Brooks, R. (2007). Eyes wide shut: The importance of eyes in infant gaze following and understanding other minds. In R. Flom, K. Lee, & D. Muir (Eds.), *Gaze following: Its development and significance* (pp. 217–241). Retrieved from http://128.95.148.60/meltzoff/pdf/07Meltzoff_Brooks_GazeChapter.pdf
Moore, C., & Corkum, V. (1994). Social Understanding at the End of the First Year of Life. *Developmental Review*. http://doi.org/10.1006/drev.1994.1014
Moore, C., & Dunham, P. J. (Eds.). (1995). *Joint Attention: Its Origins and Role in Development*. Lawrence Erlbaum.
Moors, P., Germeys, F., Pomianowska, I., & Verfaillie, K. (2015). Perceiving where another person is looking: The integration of head and body information in estimating another person's

gaze. *Frontiers in Psychology*, *6*(JUN), 1–12. http://doi.org/10.3389/fpsyg.2015.00909

Mora, K. F., & Odobez, J. (2012). Gaze estimation from multimodal Kinect data. *Gesture Recognition and Kinect Competition Workshop, CVPR*. Retrieved from http://ieeexplore.ieee.org/xpls/abs_all.jsp?arnumber=6239182

Mundy, P., & Newell, L. (2007). Attention , Joint Attention , and Social Cognition. *Current Directions in Psychological Science*, *16*(5), 269–274.

Murphy-Chutorian, E., & Trivedi, M. M. (2009). Head pose estimation in computer vision: A survey. *Pattern Analysis and Machine Intelligence, IEEE Transactions on*, *31*(4), 607–626. Retrieved from http://ieeexplore.ieee.org/xpls/abs_all.jsp?arnumber=4497208

Odobez, J., & Mora, K. F. (2013). Person Independent 3D Gaze Estimation From Remote RGB-D Cameras. *International Conference on Image Processing*. Retrieved from http://infoscience.epfl.ch/record/192423

Otsuka, Y., Mareschal, I., Calder, A. J., & Clifford, C. W. G. (2014). Dual-route model of the effect of head orientation on perceived gaze direction. *Journal of Experimental Psychology: Human Perception and Performance*, *40*(4), 1425–1439. http://doi.org/10.1037/a0036151

Parkhi, O. M., Vedaldi, A., Zisserman, A., Vedaldi, A., Lenc, K., Jaderberg, M., … others. (2015). Deep face recognition. *Proceedings of the British Machine Vision*, (Section 3).

Pelphrey, K. A., Singerman, J. D., Allison, T., & McCarthy, G. (2003). Brain activation evoked by perception of gaze shifts: The influence of context. *Neuropsychologia*, *41*(2), 156–170. http://doi.org/10.1016/S0028-3932(02)00146-X

Pérez, P., Gangnet, M., & Blake, A. (2003). Poisson image editing. *ACM Transactions on Graphics*, *22*(3), 313. http://doi.org/10.1145/882262.882269

Perrett, D., & Emery, N. J. (1994). Understanding the intentions of others from visual signals: neuropsychological evidence. *Cahiers de Psychologie Cognitive*, *13*(JANUARY 1994), 683–694.

Perrett, D. I., Hietanen, J. K., Oram, M. W., Benson, P. J., & Rolls, E. T. (1992). Organization and Functions of Cells Responsive to Faces in the Temporal Cortex. *Philosophical Transactions: Biological Sciences*, *335*(1273), 23–30.

Peterson, M. F., & Eckstein, M. P. (2013). Individual differences in eye movements during face identification reflect observer-specific optimal points of fixation. *Psychological Science*, *24*, 1216–25. http://doi.org/10.1177/0956797612471684

Recasens, A., Khosla, A., Vondrick, C., & Torralba, A. (2015). Where are they looking ? In *NIPS* (pp. 1–9).

Reid, V. M., & Striano, T. (2005). Adult gaze influences infant attention and object processing: Implications for cognitive neuroscience. *European Journal of Neuroscience*, *21*(6), 1763–1766. http://doi.org/10.1111/j.1460-9568.2005.03986.x

Saxe, R., Carey, S., & Kanwisher, N. (2004). Understanding other minds: linking developmental psychology and functional neuroimaging. *Annual Review of Psychology*, *55*, 87–124. http://doi.org/10.1146/annurev.psych.55.090902.142044

Scaife, M., & Bruner, J. S. (1975). The capacity for joint visual attention in the infant. *Nature*, *253*, 265–266.

Schweinberger, S. R., Kloth, N., & Jenkins, R. (2007). Are you looking at me? Neural correlates of gaze adaptation. *Neuroreport*, *18*(7), 693–6. http://doi.org/10.1097/WNR.0b013e3280c1e2d2

Shi, J., & Malik, J. (2000). Normalized cuts and image segmentation. *IEEE Transactions on Pattern Analysis and Machine Intelligence*, *22*(8), 888–905. Retrieved from http://doi.ieeecomputersociety.org/10.1109/CVPR.1997.609407

Simonyan, K., & Zisserman, A. (2015). Very Deep Convolutional Networks for Large-Scale Image Recognition. *Int Conf on Learning Rep*, 1–13.

Stiel, B., Clifford, C. W. G., & Mareschal, I. (2014). Adaptation to vergent and averted eye gaze. *Journal of Vision*, *14*(1), 15. http://doi.org/10.1167/14.1.15

Striano, T., & Reid, V. M. (2006). Social cognition in the first year. *Trends in Cognitive Sciences*, *10*(10), 471–476. http://doi.org/10.1016/j.tics.2006.08.006

Sugano, Y., Matsushita, Y., & Sato, Y. (2014). Learning-by-Synthesis for Appearance-Based 3D Gaze Estimation. *2014 IEEE Conference on Computer Vision and Pattern Recognition*, 1821–1828. http://doi.org/10.1109/CVPR.2014.235

Symons, L. A., Lee, K., Cedrone, C. C., & Nishimura, M. (2004). What are you looking at? Acuity for triadic eye gaze. *The Journal of General …*, *131*(4). Retrieved from http://www.ncbi.nlm.nih.gov/pmc/articles/PMC2564292/

Tanaka, M., Kamio, R., & Okutomi, M. (2012). Seamless image cloning by a closed form solution of a modified poisson problem. *Proceedings of 5th ACM SIGGRAPH Conference and Exhibition on Computer Graphics and Interactive Techniques*. http://doi.org/10.1145/2407156.2407173

Todorović, D. (2006). Geometrical basis of perception of gaze direction. *Vision Research*, *46*(21), 3549–62. http://doi.org/10.1016/j.visres.2006.04.011

Tomasello, M. (1995). Joint attention as social cognition. In C. Moore & P. Dunham (Eds.), *Joint attention: Its origins and role in developmen* (pp. 103–130). Hillsdale, NJ.

Tomasello, M. (1999). *The cultural origins of human cognition*. Harvard University Press.

Ullman, S., Harari, D., & Dorfman, N. (2012). From simple innate biases to complex visual concepts. *Proceedings of the National Academy of Sciences*, *109*(44), 18215–18220. http://doi.org/10.1073/pnas.1207690109

Vertegaal, R., Veer, G. van der, & Vons, H. (2000). Effects of gaze on multiparty mediated communication. *Graphics Interface*. Retrieved from http://www.cs.queensu.ca/home/roel/publications/2000/EffectsOfGaze.pdf

Wang, K., Wang, Y., Yin, B., & Kong, D. (2003). Face pose estimation with a knowledge-based model. In *Proceedings of 2003 International Conference on Neural Networks and Signal Processing, ICNNSP'03* (Vol. 2, pp. 1131–1134). http://doi.org/10.1109/ICNNSP.2003.1281068

Weidenbacher, U., Layher, G., Bayerl, P., & Neumann, H. (2006). Detection of Head Pose and Gaze Direction for Human-Computer Interaction. *Perception and Interactive Technologies*, *4021*, 9–19.

Wicker, B., Michel, F., Henaff, M. a, & Decety, J. (1998). Brain regions involved in the perception of gaze: a PET study. *NeuroImage*, *8*(2), 221–7. http://doi.org/10.1006/nimg.1998.0357

Wilkins, D. R. (1844). On Quaternions , or On a New System of Imaginaries in Algebra. *The London, Edinburgh and Dublin Philosophical Magazine and Journal of Science*, *25*(3), 489–495.

Wollaston, W. H. (1824). On the Apparent Direction of Eyes in a Portrait. *Philosophical Trans. Royal Soc. of London*, *114*(January), 247–256.

Wood, E., Baltrusaitis, T., Zhang, X., Sugano, Y., Robinson, P., & Bulling, A. (2015). Rendering of Eyes for Eye-Shape Registration and Gaze Estimation. In *ICCV*.

Wu, X., Kumar, V., Ross, Q. J., Ghosh, J., Yang, Q., Motoda, H., … Steinberg, D. (2008). *Top 10 algorithms in data mining*. *Knowledge and Information Systems* (Vol. 14). http://doi.org/10.1007/s10115-007-0114-2

Yovel, G., Pelc, T., & Lubetzky, I. (2010). It's all in your head: why is the body inversion effect abolished for headless bodies? *Journal of Experimental Psychology. Human Perception and Performance*, *36*(3), 759–767. http://doi.org/10.1167/9.8.460

Yu, C., & Smith, L. B. (2013). Joint attention without gaze following: human infants and their parents coordinate visual attention to objects through eye-hand coordination. *PloS One*, *8*(11), e79659. http://doi.org/10.1371/journal.pone.0079659

Zhang, X., Sugano, Y., Fritz, M., & Bulling, A. (2015). Appearance-Based Gaze Estimation in the Wild. In *CVPR* (pp. 4511–4520). Retrieved from http://arxiv.org/abs/1504.0286

## Acknowledgments

This material is based upon work supported by the Center for Brains, Minds and Machines (CBMM), funded by NSF STC award CCF-1231216, and NSF National Robotics Initiative.

## Author contributions

D.H., T.G., N.K., J.T. and S.U. designed research; D.H. and T.G. performed research; D.H. and T.G. analyzed data; D.H., T.G., N.K., J.T. and S.U. wrote the paper. D.H. and T.G. contributed equally to this work.

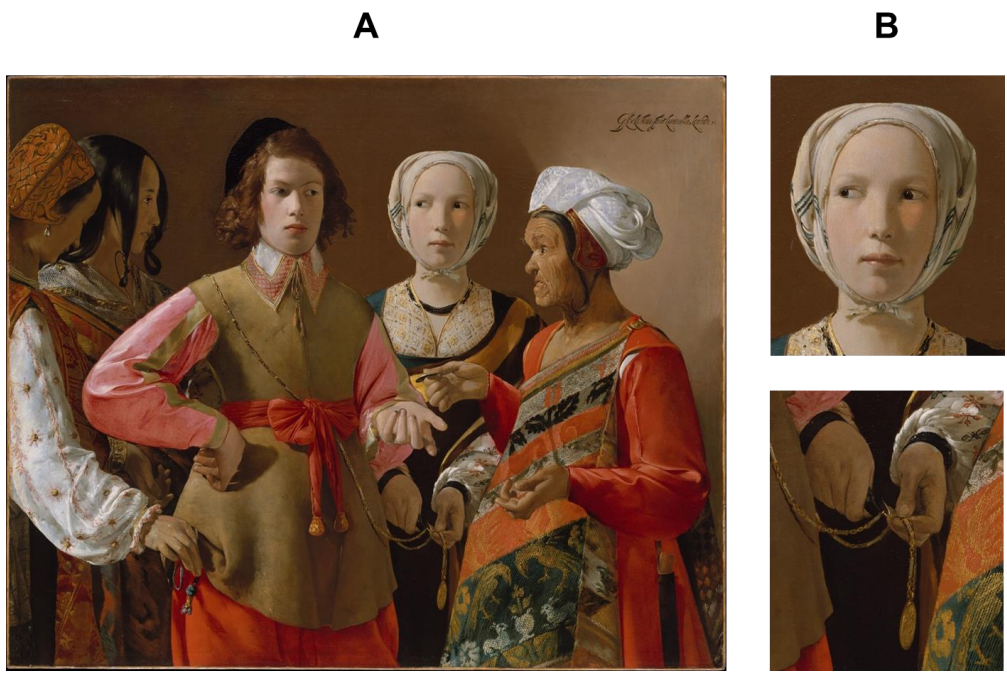

**Figure 1: Direction of gaze in social interactions. (A)** "The fortune teller", a painting by the artist Georges de La Tour (c. 1630) depicting a scene in which a young man of some wealth is having his fortune told by the old woman at right**. (B)** The young woman's eyes are averted because she is paying careful attention to the direction of gaze of the man on her right, while she cuts a medal worn by the man from its chain. The artist uses these gaze direction cues to allow the common observer a full understanding of the social act of deception and theft in the scene.

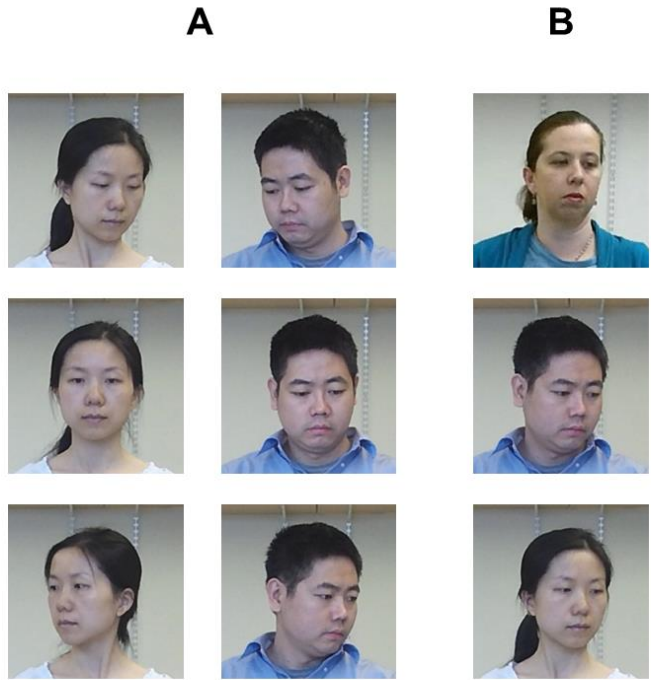

**Figure 2: Unconstraint natural looking task. (A)** Sample faces of 'lookers' extracted from the stimuli of the 'displayed' condition, illustrating the wide range of tested gaze directions and the various appearance combinations of head and eyes. **(B)** Sample faces looking at the same object in the object array, illustrating the variability of head pose and eye gaze combinations across 'lookers' for a given target.

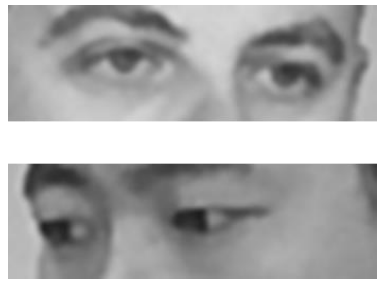

**Figure 3: The 'eyes-region'.** The region around the eyes, including the bridge of the nose and face boundaries near the temples, is sufficient for estimating direction of gaze, and essentially equivalent to the full face.

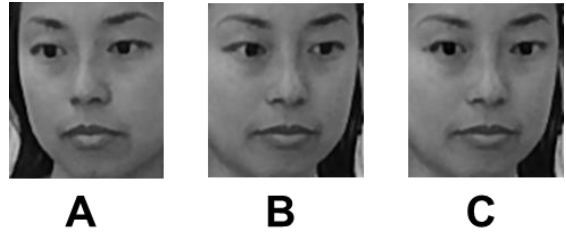

**Figure 4: The Wollaston illusion.** The face in (**A**) is perceived to gaze to the left, but when the eyes from (**A**) are extracted and combined with the head pose in image (**B**), the perceived gaze in the synthesized image (**C**), by both humans and models, is to the front. The model suggests that the perceived gaze is shifted in the direction of the new head pose. The eyes representation in these models for synthesized images like in (**C**) are different from the eyes representation for the authentic image (**A**), despite the fact that the eyes in (**A**) and (**C**) are the same. This is in agreement with Wollaston original comment that the same eyes are perceived as different with different head pose contexts.

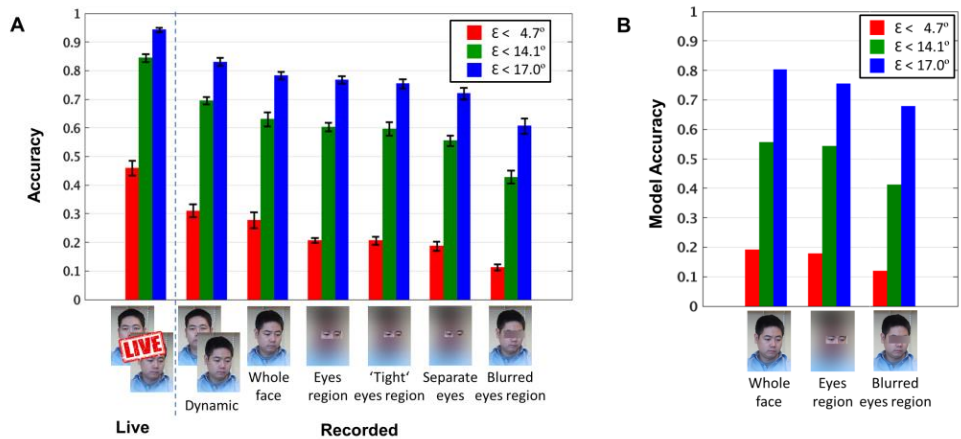

**Figure 5: Human and model accuracy under different viewing conditions.** Accuracy is measured as the percent correct of getting the exact true target object (red, angular error (ε) is less than 4.7°), or one of its surrounding neighbors (green, 4-neighborhood, ε is less than 14.1°; blue, 8-neighborhood, ε is less than 17.0°). **(A)** Human experiment results (error bars indicate standard deviation). **(B)** Results for the computational model replicate the human results under similar conditions (the dynamic condition was not tested).

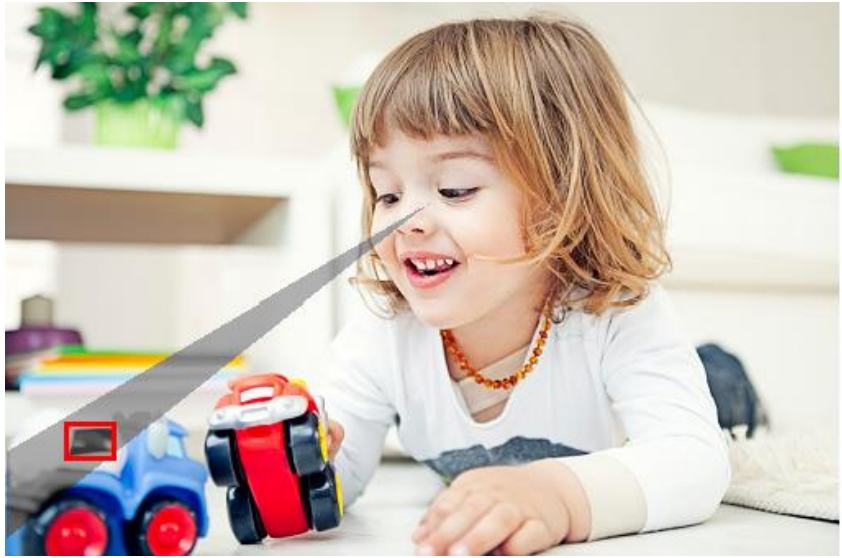

**Figure 6: Towards automatic interpretation of social interactions.** Automatic interpretation of social interactions in 2D images requires the fundamental capability of interpreting directions of gaze and detecting the attended targets. Our computational model accurately estimates the 3D direction of gaze in 2D images, and allows the detection of the attended targets by combining the 3D gaze direction with estimated depth (Liu et al., 2015) and scene segmentation (Shi & Malik, 2000).

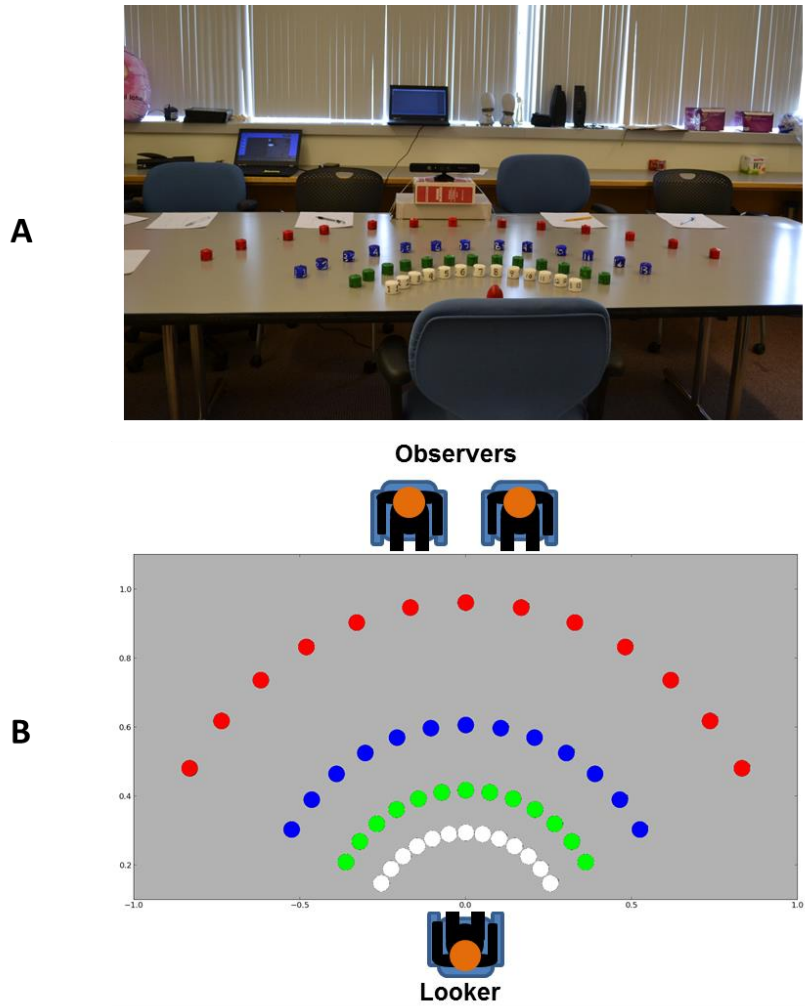

**Figure S1: Apparatus of psychophysics experiment.** An array of 52 objects (candles) was laid on a table at a concentric configuration, with 13 columns and 4 rows. A red wooden egg marked the center position of the concentric array. Objects on the same row had the same color (white, green, blue or red). The number of the column was marked on the side of each object. A laptop was placed behind the table, facing the table and the object array, to display instructing commands to the participating 'lookers'. A Microsoft Kinect V2 RGB-D sensor, positioned across the table and facing the object array, was used to record on video the looking trials of the performing 'lookers', and provide accurate 3D information of the scene for computational evaluations. (**A**) A picture of the setup taken from behind the looker's seat. (**B**) A schematic top view.

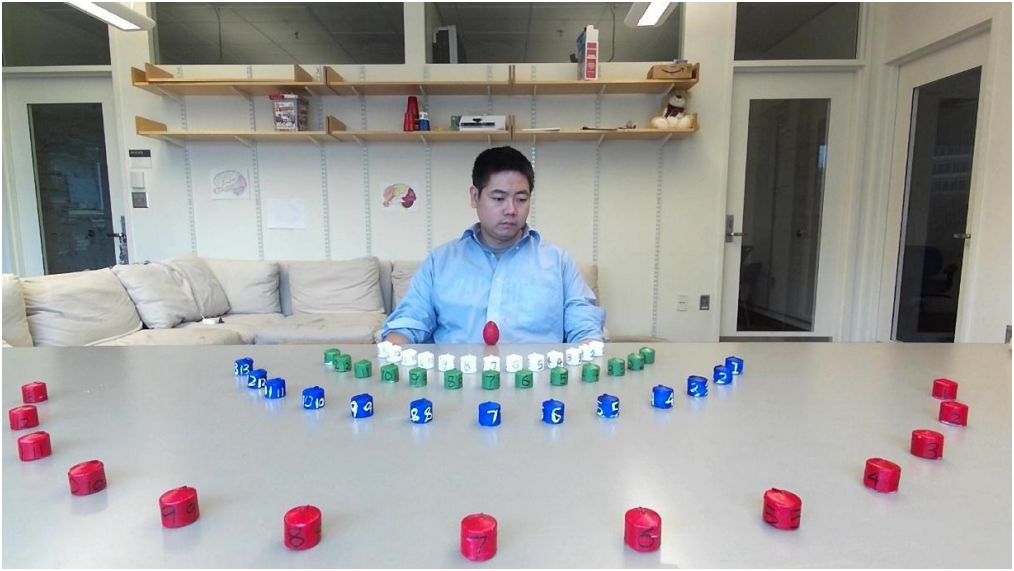

**Figure S2: Sample 'recorded' static stimulus.** Visual stimuli were displayed on a screen, depicting a 'looker' freely looking at one of 52 targets on the table. In each trial, the 'observer' was asked to mark on the displayed image his best guess of the looker's target using a computer mouse.

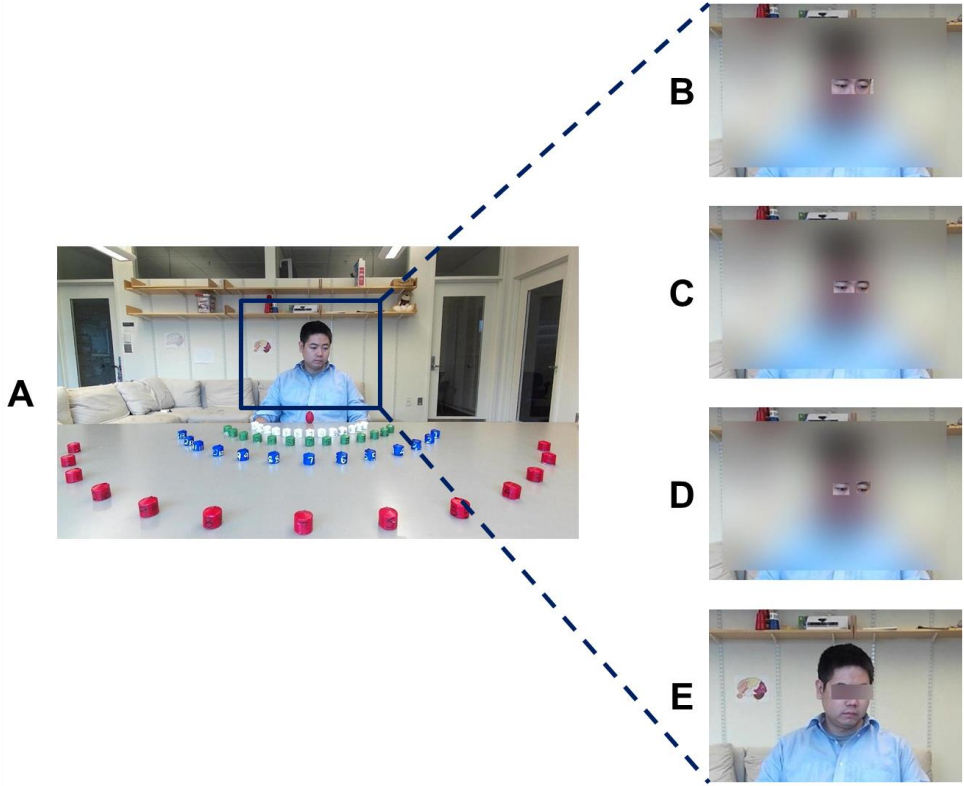

**Figure S3: Static 'displayed' conditions.** Static 'displayed' stimuli consisted of 5 viewing conditions: (**A**) whole face visible; (**B**) 'eyes-strip' only; (**C**) tight 'eyes-strip' only; (**D**) separate eyes only; (**E**) face without eyes.

| Input image | Gaze direction | | | | Head orientation |
|---|---|---|---|---|---|
| | 'Exact' accuracy | '4-neighbors' accuracy | '8-neighbors' accuracy | Mean angular error | Mean angular error |
| Full eyes-region | 13.15% | 46.14% | 67.80% | 9.58º | 10.22º |
| Eyes-region w/o nose bridge | 11.12% | 40.87% | 61.01% | 10.44º | 11.27º |
| Eyes-region w/o face boundaries | 5.01% | 21.66% | 35.23% | 16.38º | 13.93º |
| Eyes-region w/o both nose bridge and face boundaries | 4.23% | 17.69% | 29.75% | 18.48º | 15.73º |

**Table S1: Informative cues in the eyes-region.** Performance of a computational model trained to estimate both head orientation and direction of gaze from images of the eyes-region (single-stage, DNN), when applied to images from the computational evaluation dataset, under different occlusion conditions. The results indicate that the information is originated not only in the eyes, but also in the nose bridge and face boundaries, which contribute information mainly for head orientation.

| Model | | | Gaze direction | | | | Head orientation |
|---|---|---|---|---|---|---|---|
| | | | 'Exact' accuracy | '4-neighbors' accuracy | '8-neighbors' accuracy | Mean angular error | Mean angular error |
| Single-stage | Whole-face | *k*-NN | 13.94% | 45.09% | 69.94% | 9.40º | 6.81º |
| | | DNN | 8.25% | 36.85% | 62.47% | 10.58º | 9.55º |
| | Eyes-region | *k*-NN | 13.31% | 45.72% | 65.76% | 9.64º | 9.07º |
| | | DNN | 13.15% | 46.14% | 67.80% | 9.58º | 10.22º |
| Two-stage | Whole-face | *k*-NN | **19.15%** | **54.70%** | **75.26%** | **8.39º** | **6.81º** |
| | Eyes-region | *k*-NN | 15.55% | 51.10% | 73.64% | 8.73º | 9.07º |

**Table S2: Performance evaluation of the computational models.** A comparison of estimated target accuracy, and mean angular error of estimated direction of gaze and head orientation, between the two-stage computational model and the alternative single-stage models. The results are reported for the computational evaluation dataset.

## Supplementary material

*'Recorded' stimuli*

All visual stimuli in experiments 2a and 2b were acquired using A Microsoft Kinect V2 RGB-D sensor. Still images were extracted as individual frames from the recorded videos. The pixel resolution, for both video and image stimuli, was 1920×1080. There were 6 different viewing conditions:

(*RC1*) Dynamic videos. Video clips, 10 seconds long for each trial, depicting a performing looker. Each video clip starts when a looker is at a resting state pose (watching a command on the screen behind the recording system), then showing the full motion trajectory of the looker until the looker fixates her gaze at a target on the table. The video clip ends when the looker returns to the resting state pose (waiting for the next trial).

(*RC2*) Whole-face still images. Image frames of lookers gazing at targets (one image for each target) were extracted from the recorded videos, when the lookers keep steady both their head and eyes while fixating on a target. All image parts are clearly visible, including the looker's face and the object array on the table (Fig. S3A). Faces are of width 86±8 pixels and height 97±7 pixels.

(*RC3*) Eyes-region still images. The images from *RC2* were manipulated as follows: a rectangular image region around the eyes is kept clearly visible, including the nose bridge and face boundaries at the temples (width 58±4 pixels; height 21±3 pixels). A surrounding image region of size 300×200 pixels, including the rest of the face, is blurred, by down-sampling followed by up-sampling of the image region by a factor of 50, using the Lanczos resampling method (Fig. S3B).

(*RC4*) Head-only still images. The images from *RC2* were manipulated as follows: a rectangular image region around the eyes (the same region from *RC3*) is blurred (using the same blur method described above for *RC3*), while the rest of the face is kept clearly visible (Fig. S3E).

(*RC5*) Tight-eyes still images. The images from *RC2* were manipulated as follows: a rectangular image region tightly surrounding the eyes is kept clearly visible, including the nose bridge but excluding any face boundaries (width 49±4 pixels; height 15±2 pixels), while a surrounding image region of size 300×200 pixels, including the rest of the face, is blurred (using the same blur method described for *RC3*, Fig. S3C).

(*RC6*) Separate-eyes still images. The images from *RC2* were manipulated as follows: two rectangular image regions tightly surrounding each of the eyes is kept clearly visible (left eye region width 24±8 pixels, height 18±5 pixels; right eye region width 20±3 pixels, height 16±2 pixels), while a surrounding image region of size 300×200 pixels, including the rest of the face (in particular, including the nose bridge and face boundaries), is blurred (using the same blur method described for *RC3*, Fig. S3D).

For generating the stimuli in *RC3-RC6*, image center locations of the eyes, nose and mouth, as well as the face bounding box, were automatically detected by the Kinect's face tracking algorithm. The location and dimensions of the rectangular image regions around the eyes were heuristically determined by the locations of the face parts above and their spatial configuration, in every image.

*Two-stage computational model*

We developed a model for estimating the 3D direction of gaze from 2D images of faces, using computer vision methods. During training, the model learns to associate facial appearances (either whole-face or eyes-regions) with 3D head pose directions, and to associate eye-pairs appearances with 3D offset directions between head orientation and final gaze direction. In our implementation we used HOG image descriptors (2480 dimensions) to represent facial appearances, and dense-SIFT image descriptors (8448 dimensions) to represent eye-pairs appearances (Bosch et al., 2006; Dalal & Triggs, 2005; Lowe, 2004). During inference, the processing in the model works in two stages. In the first stage, the model estimates the 3D head pose direction from the input face image, by extracting familiar faces in the model with similar facial appearances (based on *k*-nearest neighbors technique, *k*=15, (Duda et al., 2001; Wu et al., 2008)). The estimated head orientation is the weight-average (based on appearance similarity) of the associated head pose directions with the neighboring faces. In the second stage, the model extracts familiar faces in the model with similar eye-pairs appearances, only around the estimated head orientation from the first stage. The associated 3D offset directions between the head orientation and the final gaze direction of the extracted neighboring faces, are weight-averaged to yield the estimated 3D offset direction from the estimated head orientation from the first stage and the final gaze direction. In our implementation we use rotational quaternions to represent 3D directions (Altmann, 1986; Wilkins, 1844).

*Alternative computational models*

We compared the performance of our two-stage model with alternative appearance-based models that infer directly head orientation and final direction of gaze from the input image, in a single processing stage. We trained and tested additional nearest neighbor models and deep neural network models, which all yielded inferior performance compared with the two-stage model (Table S2).

The nearest neighbor models included models associating HOG or dense-SIFT appearance descriptors of the whole face or the eyes-region directly with head orientation and gaze direction. The deep neural network models were based on face recognition models (Parkhi et al., 2015), fine-tuned to minimize the error between the estimated and ground-truth head orientation and gaze direction, given input images of the whole face or the eyes-region (the original '*softmax*' output classification layers were replaced with '*fully connected*' output layers of size 4, representing 3D directions in rotational quaternions).